\documentstyle[twocolumn,aps,prb,epsfig,floats]{revtex}

\begin{document}
\twocolumn[\hsize\textwidth\columnwidth\hsize\csname@twocolumnfalse\endcsname

\title{Equilibrium valleys in spin glasses at low temperature}

\author{E. Marinari$^{(1)}$,
O. C. Martin$^{(2)}$,
F. Zuliani$^{(2)}$}

\address{
1) Dipartimento di Fisica, INFM and INFN,  \\
Universit\`a di Roma {\em La Sapienza},  \\
P. A. Moro 2, 00185 Rome, Italy\\
2) Laboratoire de Physique Th\'eorique et Mod\`eles Statistiques, \\
b\^at. 100, Universit\'e Paris-Sud, F--91405 Orsay, France}

\date{\today}
\maketitle

\begin{abstract}
We investigate the $3$-dimensional Edwards-Anderson spin glass model
at low temperature on simple cubic lattices of sizes up to $L=12$. Our
findings show a strong continuity among $T>0$ physical features and
those found previously at $T=0$, leading to a scenario with emerging
mean field like characteristics that are enhanced in the large volume
limit. For instance, the picture of space filling sponges seems to
survive in the
large volume limit at $T>0$, while entropic effects play a crucial
role in determining the free-energy degeneracy of our {\em finite
volume states}. All of our analysis is applied to equilibrium
configurations obtained by a parallel tempering on $512$ different
disorder realizations.  First, we consider the spatial properties of
the sites where pairs of independent spin configurations differ and we
introduce a modified spin overlap distribution which exhibits a
non-trivial limit for large $L$. Second, after removing the
$Z_2$ ($\pm 1$) symmetry, we cluster spin configurations into {\em
valleys}. On average these valleys have {\em free-energy} differences
of $O(1)$, but a difference in the (extensive) {\em internal energy}
that grows significantly with $L$; there is thus a large interplay
between energy and entropy fluctuations. We also find that valleys
typically differ by sponge-like space filling clusters, just as
found previously for low-energy system-size excitations above the
ground state.
\end{abstract}


\twocolumn]\narrowtext

\section{Introduction}
\label{sect_intro}

In the past two decades there has been a large amount of theoretical,
experimental, and numerical work on spin glasses (see for example
\cite{MezardParisi87b,Young98}, and references therein).  Two
different points of view have dominated the frameworks used by most
researchers. On one side, there is the idea that mean field
features~\cite{Parisi79,Parisi79b,Parisi80,Parisi80b,Parisi80c,Parisi83},
that are astonishingly different from those encountered in systems
without frustration, are relevant in realistic, finite dimensional
spin glasses. On the other side, there is the expectation that the
(non-trivial) extension~\cite{BrayMoore86,FisherHuse88} of scaling
ideas to these frustrated systems should be the correct framework for
short range spin glasses. The results contained in this note will give
support to the point of view that many features of the finite
dimensional Edwards-Anderson (EA) \cite{EdwardsAnderson75} spin
glasses are common to the ones of the mean field Parisi solution and
are well described in terms of the Replica Symmetry Breaking (RSB)
approach.  This point of view is already supported by a large number
of numerical simulations and by studies of ground states. Indeed,
numerical simulations indicate, among other features, that the EA
model has an average overlap probability distribution $P(q)$ which is
broad in the thermodynamic limit, signaling (continuous) RSB (see
\cite{MarinariParisi99b} and references therein). Similarly, ground
state computations
\cite{KrzakalaMartin00a,PalassiniYoung00a,MarinariParisi00a,MarinariParisi00b,HedHartmann00a,HoudayerKrzakala00}
show that the energy landscape of the EA model is characterized by
system-size valleys whose excitation energies remain of order one in
the infinite volume limit. The consequence of this property on the
system at temperature $T>0$ is likely to be RSB. This is typically
expected to imply that for a given instance of the disorder (one given
realization of the quenched random couplings) spin configurations will
cluster into multiple ``valleys'', not related by the up-down symmetry
(there is no magnetic field so we shall always work modulo the $\pm 1$
symmetry).

The purpose of this work is to study the $3$-d EA model at low
temperature, and to characterize its valleys, both thermodynamically
and for their position-space (i.e., spatial) properties. Among other
things, we ask whether valleys having $O(1)$ free-energy differences
have similar energies and entropies; if not, RSB implies that the
fluctuations in these quantities must nearly cancel. Also, what are
the typical differences among valleys?  How does one go from one
valley to another?  Before addressing these questions, we have to be
far more specific in defining a valley. Our definition is algorithmic,
based on a method for clustering configurations according to their
spin overlaps (or equivalently according to their Hamming
distances). Such a clustering scheme has been used successfully by Hed
{\it et al.}~\cite{HedHartmann00a} for analyzing spin glass ground
state configurations.  By applying such a clustering algorithm, we
determine at a given temperature the ``valleys'' that characterize
each disorder realization.  Then we compute various statistical
properties of these valleys and of their associated TAP-like states.
Interestingly, the picture obtained from looking at $O(1)$ energy
system-size excitations above the ground state
\cite{HoudayerKrzakala00} also applies to our finite temperature
valleys. One goes from one valley to another by flipping system-size
clusters of spins that are space filling and topologically highly
non-trivial: this justifies calling them {\em spongy
clusters}~\cite{HoudayerMartin00b}.

The outline of this paper is as follows.  First we specify how our
equilibrium configurations were generated.  Then, given two such
configurations, we consider various ways to compare them with a
special attention on topological properties of the set of spins where
they differ. We construct the clusters of these spins and
classify them into {\em sponge-like}, {\em droplet-like}, or neither
of these.  Our data indicate that, if the statistics of such
topological events are monotonic with $L$, sponge-like events arise
with a strictly positive probability in the large $L$ limit. Then we
describe how we define the {\em valleys} into which we will cluster
our equilibrium configurations (we give additional details about the
clustering algorithm in the appendix).  Not surprisingly,
configurations assigned to significantly different valleys nearly
always have sponge-like differences, i.e., they do not differ in
localized, droplet like areas, but in a diffuse set of sites. We
compute the (extensive) internal energy of each valley: we find that
this quantity fluctuates dramatically from valley to valley, growing
very clearly with the lattice size $L$.  Finally we consider the
properties of the TAP-like states associated with the valleys.  We
find that again the difference among different TAP-like states is
typically given by a spongy cluster, just as in the zero temperature
case. We also study the link overlap distribution among these states.

\section{The model and its numerical simulation}
\label{sect_model}

We consider the $3D$  Edwards-Anderson (EA) spin glass
model, whose Hamiltonian is

\begin{equation}
H = - \sum_{\langle ij\rangle} J_{ij} S_i S_j \ ,
\end{equation}
where $\langle ij\rangle$ indicates that the sum is over nearest
neighbors.  We work on simple cubic lattices with pe\-rio\-dic
bo\-un\-da\-ry conditions and zero magnetic field. The quenched
couplings $J_{ij}$ are independent random variables that can take the
two values $\pm 1$ with equal probability $1/2$. In this study we
have used spin configurations at $T=0.5$. (Note that
$T_c \simeq 1.1$.) We will
present data taken on lattices of linear size $L=6$, $8$ and $12$.

We rely on a parallel tempering updating scheme (see for example
\cite{Marinari98} and references therein) that is very effective for
simulating EA spin glasses. The different temperatures used in the
runs are uniformly spaced. We use $19$ temperature values going from
$T_{min}=0.5$ to $T_{MAX}=2.3$ on the $L=6$ lattices, $49$ temperature
values going from $T_{min}=0.5$ to $T_{MAX}=2.1$ on the $L=8$
lattices, and $64$ temperature values going from $T_{min}=0.5$ to
$T_{MAX}=2.075$ on the $L=12$ lattices. We have checked in detail that
we reach a very good thermalization according to all the usual
tests~\cite{Marinari98}: the acceptance ratio of the tempering sweeps
is high (always of order $0.7$), the permanence histograms are very
flat, and observables look very constant when plotted as a function of
the log of the elapsed Monte Carlo time. In addition, we have checked
that on individual disorder instances the $\pm 1$ symmetry of $P_J(q)$
is very well satisfied, see for example figures~\ref{fig_P_J_TWO}
and~\ref{fig_P_J_THREE}. In short, since we have used very long MC
runs with very conservative and safe choices of the parameters,
thermalization seems to be complete.

For each disorder instance and lattice size, we perform the parallel
tempering {\it independently} on two (sets of) replicas.  For each, we
first run $10^6$ MC sweeps just to reach thermalization; a sweep
consists of $L^3$ trial spin flips and one tempering $\beta$ trial
update.  Following that, we run for $1.1\cdot 10^6$ MC sweeps to
gather (equilibrium) statistics.  During these sweeps we save
$1100$ spin configurations, that is one every $1000$ MC sweeps, giving
a total of $2200$ equilibrium configurations for that disorder
sample. These are the configurations that are used for the analysis
described in the rest of this paper. The runs to generate these
configurations took on the order of $2$ weeks of machine time on a
dedicated Linux cluster based on $16$ Pentium II $450 MHz$ processors.

For each lattice size ($L=6$, $8$, and $12$), we thus have $16 \times
32 = 512$ different disorder samples (the factor $32$ comes from the
use of multi-spin coding on each of the $16$ machines), and a total of
$2 \times 1100$ equilibrium spin configurations for each
disorder sample.

\section{Droplet and sponge differences of equilibrium configurations}
\label{sect_topology}

Each spin configuration $C$ appears at equilibrium, at temperature
$T$, with probability $P_B(C) = Z_J^{-1} \exp(\frac{-H(C)}{T})$ (where
$Z_J$ is the partition function which normalizes the Boltzmann weight
for that instance).  Consider a pair of spin configurations $C^{(1)}$
and $C^{(2)}$, selected independently according to their \`a priori
probability, and let $G$ be the set of sites where the spins of the
two configurations differ. One defines the spin overlap by

\begin{equation}
  q(C^{(1)},C^{(2)}) = \frac{1}{N} \sum_{i=1}^N S_i^{(1)} S_i^{(2)}\ ,
\end{equation}
where in our $3$-d case $N=L^3$. If $|G|$ is the cardinality of $G$,
then $q = 1-2|G| / N$.  In figure~\ref{fig_P_J_TWO} we show an example
(for $L=12$) of a distribution $P_J(q)$ of overlaps for one
realization of the disorder (the dotted curve), and with the solid
curve we show the distribution averaged over the disorder, that is
over our $512$ disorder samples. Because of the up-down symmetry, in
the rest of this work we will only look at the $q\ge 0$ part of the
support, and we will ignore the part for negative $q$ values.  For the
disorder instance used in figure~\ref{fig_P_J_TWO}, most of the
overlaps are close to the two values associated with the positions of
the two peaks of $P_J(q)$. This suggests that in this instance the
spin configurations fall into two ``valleys'', as will be confirmed
later. In figure~\ref{fig_P_J_THREE} we consider a different disorder
instance where $P_J(q)$ has three peaks (always in the $q>0$
sector). Because of this kind of behavior, we should investigate what happens
when we cluster the configurations according to their mutual overlaps;
however, before doing so, let us first discuss how two equilibrium
configurations differ {\it spatially}.

\begin{figure}
\centering
\epsfig{file=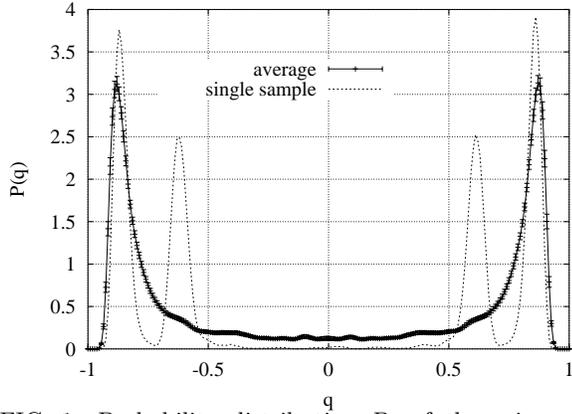,width=0.45\textwidth}
\caption{Probability distribution $P_J$ of the spin overlap 
$q(C^{(1)}, C^{(2)})$ for a given disorder instance (dotted curve), and its
disorder average $P$ (solid curve).}
\label{fig_P_J_TWO}
\end{figure}

\begin{figure}
\centering
\epsfig{file=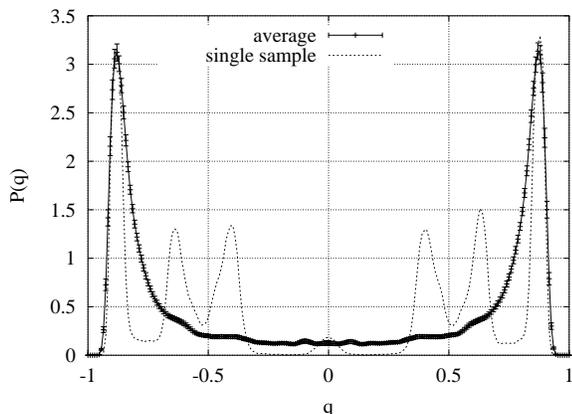,width=0.45\textwidth}
\caption{As in figure \ref{fig_P_J_TWO}, but for a disorder instance 
that generates three peaks in $P_J$ at $q>0$.}
\label{fig_P_J_THREE}
\end{figure}

When two configurations are in the same valley, we expect them to
differ only by droplet-like objects; the set of spins $G$ forming
their difference should consist of small clusters.  On the contrary,
when considering configurations in two different valleys, we expect
$G$ to contain a large and topologically non-trivial cluster. To
quantify this, we follow reference \cite{HoudayerKrzakala00} and
declare $G$ to be {\em sponge-like} if it {\bf and} its complement
wind around the lattice in all three directions, whereas we call it
{\em droplet-like} if it does not wind around any of the three directions
of the lattice.  Finally we label {\em intermediate} the other cases
of windings.  Our data show very clearly a strong correlation between
the (relative) size of $G$ and its topological class: the larger
${|G|}/{L^3}$, the more likely it is to be sponge-like. Do sponge-like
differences between equilibrium configurations survive in the large
$L$ limit? In table~\ref{tab_typeOfG} we give the probabilities of the
three topological classes as a function of the lattice size. Just as
in the $T=0$ energy landscape studies \cite{HoudayerKrzakala00}, the
frequencies of sponge-like events {\it increases} as $L$ increases,
suggesting that in the limit $L \to \infty$ equilibrium configurations
differ by sponge-like clusters with a strictly positive probability.

\begin{table}
\begin{center}
\caption{Probability of each of the $3$ topological classes for
differences of independent spin configurations.}
\label{tab_typeOfG}
\begin{tabular}{llll}
\hline\noalign{\smallskip}
L & sponge-like & intermediate & droplet-like \\
\noalign{\smallskip}\hline\noalign{\smallskip}
6  & $ 0.082 \pm 0.005 $
   & $ 0.157 \pm 0.007 $
   & $ 0.761 \pm 0.010 $
   \\
8  & $ 0.104 \pm 0.006 $
   & $ 0.166 \pm 0.007 $
   & $ 0.730 \pm 0.010 $
   \\
12 & $ 0.117 \pm 0.007 $
   & $ 0.160 \pm 0.006 $
   & $ 0.723 \pm 0.010 $                      
   \\                              
\noalign{\smallskip}\hline
\end{tabular}
\end{center}
\end{table}

Coming back to the distribution of overlaps, let us decompose $P(q)$
into the sum of three densities, one for each topological class:

\begin{equation}
  P(q) = \rho_s (q) + \rho_d (q) + \rho_i (q)\ ,
\end{equation}
where $\rho_s$ is the sponge density, $\rho_d$ the droplet density,
and $\rho_i$ the intermediate density. (A different decomposition
based on valleys has been introduced by Hed {\it et
al.}~\cite{HedHartmann00b}; our method has the great advantage that it
does not rely on any clustering into valleys and is thus parameter-free.) Now consider only those
pairs of configurations where $G$ is sponge-like: we show in
figure~\ref{fig_rho_s} the density distribution $\rho_s$ of the
overlaps belonging to this topological class.  The main trend is the
widening of the distribution's support: sponge-like $G$'s can occur
for increasing values of $q$ as $L$ grows.  Other than that, for small
$q$, $\rho_s$ is stable when $L$ increases from $6$ to $12$: this
constitutes further numerical evidence for replica symmetry breaking
in the three-dimensional EA model.

\begin{figure}
\centering
\epsfig{file=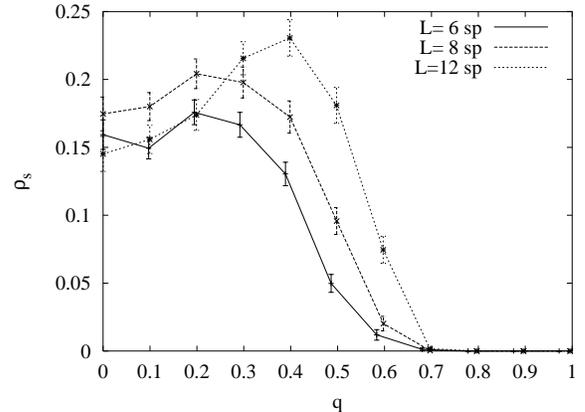,width=0.45\textwidth}
\caption{The density distribution $\rho_s$. This is the restriction of
$P(q)$ to overlaps associated with configurations $C^{(1)}$ $C^{(2)}$
having sponge-like differences.}
\label{fig_rho_s}
\end{figure}

\section{Clustering equilibrium configurations}
\label{sect_clustering}

We want to give an appropriate definition of a {\em valley}.  One
would like a valley to be a connected region in phase space that
contributes a finite measure to the partition function (this statement
can be made precise by coarse graining the phase
space). Configurations belonging to the same valley are then expected
to be close (i.e., they have a large overlap), while configurations
belonging to different valleys are expected to be far (small
overlap). Now, in more precise terms, given a collection of
equilibrium configurations, we seek to cluster them into families
which hereafter we shall refer to as {\em valleys}.  Although this
kind of classification problem is generally ill-posed, one expects
good clustering methods to lead to nearly identical results when the
valleys are {\it well} separated. Such an ``ideal'' case arises in
models with one-step RSB: there, the overlap between two equilibrium
configurations converges in the thermodynamic limit to one of two
possible values, so assigning configurations to clusters is
straight-forward.

In the case of {\em continuous} RSB, the situation is slightly
ambiguous because in the thermodynamic limit we expect that there will
be some hierarchical organization of the states: there will be valleys
inside valleys inside valleys...  Nevertheless, it is still meaningful
to define valleys, as long as the splitting of valleys into
sub-valleys is not relevant for the observables considered.  If that
is indeed the case, one can in fact allow for valley sub-divisions ad
infinitum as was done by Hed {\it et al.}~\cite{HedHartmann00a}: those
authors clustered ground state configurations in the $3$-d $\pm J$ EA
model, letting the sub-valley sizes become arbitrarily small.  Since
our goal here is to focus on well separated valleys, we construct our
valleys up to some cut-off overlap value $q^*$; in effect, all
sub-valleys with overlaps larger than $q^*$ are lumped together.
After, we will check by using different values of $q^*$ that our
findings are not sensitive to that cut-off; this can be so of course
only when considering observables associated with overlaps smaller
than that cut-off.

For completeness, we give here a high-level description of our
clustering method. The appendix gives some details about the algorithm
itself (a similar approach was defined by Iba and
Hukushima~\cite{IbaHukushima00}).  Conceptually, our assignment of
equilibrium configurations to 
a given valley is motivated by the TAP states\cite{ThoulessAnderson77} 
arising in the mean field picture.  Each of
our valleys will be associated with a ``TAP-like'' state which for
our purposes is just a set of magnetizations, one for each site.
In analogy with TAP states, these magnetizations are the thermal
averages of the spin variables, restricted to configurations belonging
to the valley of interest; this is like using a
``finite volume equilibrium state'' to define the TAP states. Thus, given
the set of
configurations assigned to a valley, we compute the magnetization
$m_i$ at site $i$ by finding the mean of the values of the $S_i$ over these
configurations.  If we want to have $M$ valleys, we are to find $M$
TAP-like states; algorithmically, we find these $M$ states and their
associated valleys by iteration.  The self-consistency conditions are
that:

\begin{enumerate}

\item the TAP-like state of valley $k$ is defined by the
magnetizations $m_i$ of that valley;

\item a configuration belongs to the valley $k$ if the TAP-like state
nearest to it is number $k$ (the nearness is measured by the overlap)
{\em and if} that overlap is greater than $q^*$.

\end{enumerate}

If $q^*$ is close to $1$ only very similar configurations will be
included into the same valley; on the contrary, if $q^*$ is small most
of the time all the configurations will be assigned to just one
valley.  Intuitively, one can think of $q^*$ as being a resolution; as
one increases $q^*$, valleys get resolved into sub-valleys, and as
$q^* \to 1$, one reaches the limit where each valley corresponds
to a single spin configuration.

Our self-consistent procedure does not generally determine the
clustering uniquely, but as was said before, any reasonable clustering
algorithm should do well if the valleys are sufficiently separated
(that is the case we want to consider in this work). In practice we
find the valleys (and their TAP-like states) iteratively: we start by
defining a first valley, then when possible and needed we add and
define a second valley, and so on. For a given value of the cut-off
$q^*$, one clusterizes more and more of the configurations when
increasing the number of TAP-like states. One stops the construction
when any of the following conditions are met: (i) all spin
configurations have been assigned to valleys; (ii) a maximum number of
valleys (namely ten in our code) has been reached; (iii) increasing
the number of TAP-like states would lead to too similar valleys, that
is some of the overlaps of the TAP-like states would become greater
than $q^*$.

At the end of the procedure, the algorithm gives
us valleys (families of configurations) and their associated TAP-like
states. Not surprisingly, for instances where the clustering leads to
several sizeable valleys, the $P_J(q)$ has several peaks, and the
reverse is also true. Furthermore, we checked the property
previously mentioned in section~\ref{sect_topology}, namely 
that configurations in valleys with small mutual overlaps usually
differ by sponge-like clusters.

To give the reader a bit of intuition about how the algorithm behaves,
let us show what happens for the disorder instance used
in figure~\ref{fig_P_J_TWO}. For that instance, there
are two peaks in $P(q)$, one near $q=0.6$ and the other near 
$q=0.8$. Thus we expect to find two valleys with self-overlap
close to $0.8$ and whose mutual overlap should be near
$0.6$. Recall that $q^*$ acts as a resolution; let us then follow
the result of the clustering as we go from $q^*=0$ (low
resolution) to $q^*=1$ (high resolution).

For small
values of $q^*$, the algorithm puts nearly all the configurations
into one huge valley. (Recall that we work modulo
the $\pm 1$ symmetry; the Hamiltonian has no magnetic field and 
our clustering procedure respects that symmetry by always
working modulo this symmetry. In effect, each configuration
can be identified with a pair of opposite configurations.) The 
self-overlap of the corresponding TAP-like state
is quite stable, going from $0.741$ at $q^*=0$ to
$0.752$ at $q^* = 0.61$. Then at $q^*=0.62$,
this big valley breaks up into two sub-valleys. The break-up is
accompanied by a discontinuity in 
the self-overlap: for the largest valley,
it jumps to $0.86$.  Also, the overlap between the two TAP-like
states is $0.61$. This value is no surprise 
as the algorithm forbids TAP-like states
with overlaps greater than $q^*$; when
new valleys appear, one of the inter-valley overlaps should correspond
precisely to the cut-off. 

Now we increase $q^*$ further. For the range $0.62 \le q^* \le 0.72$,
the assignment of the $2200$ configurations to the $2$ valleys is
$q^*$-independent. Beyond $q^*=0.72$, things are qualitatively similar
except that a small number of configurations
leave these valleys. This can be
understood quite simply: the configurations in the valley that are the 
furthest
from the TAP-like center will be pushed out first. Finally, when 
$q^* > 0.82$, the largest valley breaks up and we end up with more small
valleys. Clearly, once $q^*$ is close to $q_{EA}$, one will not keep a
small number of valleys; such values of $q^*$ should not be
used. Even in the case of systems without frustration and
disorder, this would lead to multiple valleys.

\section{Valley-to-valley energy fluctuations}
\label{sect_E_fluctuations}

The main goal of this work is to understand how valleys differ; to
this end it is convenient, among the $512$ disorder instances we have
produced, to focus only on those that give rise to more than one
valley.  For each instance of that type, the $2200$ equilibrium
configurations cluster into two or more valleys as generated by our
algorithm. (For the data shown hereafter, the resolution is $q^*=0.5$.) However,
sometimes one of these valleys will be of very small size, containing
$1$ or $2$ configurations. Since we want to investigate {\em typical}
properties, we have decided to study only valleys of an acceptable
size.  Quantitatively, we do this by restricting ourselves to the two
largest valleys and by demanding that each of them include a
substantial fraction of the $2200$ spin configurations.
 
The analysis we present in this section has been performed using those
disorder instances where the two largest valleys together contain at
least $80\%$ of the configurations, and the second largest valley at
least $20\%$. Out of the $512$ disorder instances, these selection
criteria left us with $81$ instances for $L=6$, $91$ instances for
$L=8$, and $91$ instances for $L=12$. We have repeated the analysis
with different thresholds and have found that our conclusions are
unaffected by the change. We now begin by discussing the thermodynamic
properties of these valleys, while in the sections thereafter we will
focus on the corresponding TAP-like states.

The free-energy $F_r$ of a valley $r$ is defined as

\begin{equation}
  e^{- F_r / T } = \sum{'}~ e^{- H(C) / T }\ ,
\end{equation}
where $\sum{'}$ indicates that only those configurations $C$ belonging
to valley $r$ are to be included. Note that this sum is the {\it
weight} of that valley, equal to the valley's contribution to the
partition function $Z_J$.  Since the Monte Carlo samples
configurations according to their Boltzmann factor, it visits the
different valleys according to their contribution to $Z_J$. Thus, the
{\it number} of times the Monte Carlo generates a configuration in
valley $r$ is proportional to $\exp (-F_r / T)$.

For convenience, we order our valleys according to their size (the
largest valley includes $|V_1|$ configurations, the next largest
$|V_2|$, and so on).  As a result, the valleys are ordered by
increasing free-energies.  Defining $\Delta F = F_2 - F_1 \ge 0$, we
can estimate this difference from our data as

\begin{equation}
  \exp \left[ - \frac{\Delta F }{ T} \right] = \frac{|V_2| }{ |V_1|}\ .
\end{equation}
In the mean field picture, $\Delta F = O(1)$ with a finite probability
at large $L$. It is not difficult to see that our selection criteria
impose $\Delta F = O(1)$; what is then relevant is the fraction of
disorder instances that pass these selection criteria. These fractions
are $15\%$, $17\%$, and $17\%$ for $L=6$, $8$, and $12$. Within the
error bars, these fractions are stable, in agreement with the mean
field picture.

\begin{figure}
\centering
\epsfig{file=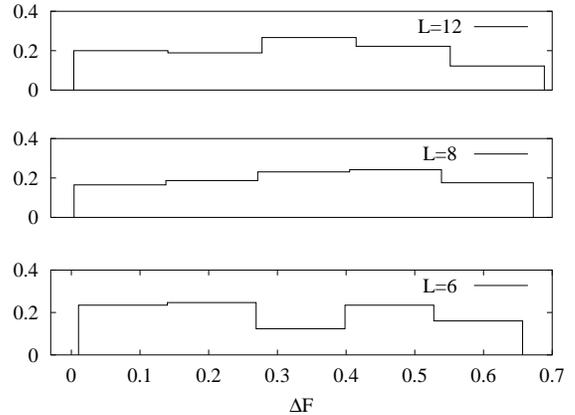,width=0.45\textwidth}
 \caption{Probability distributions of $\Delta F$, for $L=6$, $8$, and $12$.}
\label{fig_P_DeltaF}
\end{figure}

Another prediction of the mean field theory is that $\Delta F$ has an
exponential distribution, with a slope becoming small as $q^*$
decreases. In figure~\ref{fig_P_DeltaF} we show the numerical values
we have found for the probability distribution of $\Delta F$ at $L=6$,
$8$, and $12$. The distribution seems quite insensitive to $L$ and its
shape is rather flat.  Again we interpret the data as being compatible
with the mean field picture.

Each valley can be thought of as a finite volume equilibrium state;
intensive observables are expected to have the same values across
different valleys, but extensive quantities can very well fluctuate
significantly.  Let us thus consider the (extensive) internal energy
of valley $r$: $E_r$ is defined as the average of the energies of the
configurations belonging to that valley. We have measured the
statistical properties of $\Delta E \equiv E_2 - E_1$.  First, we find
that the mean of this random variable is positive; that can be
understood by saying that since by construction $\Delta F > 0$, all
other things being equal, one expects $\Delta E >0$.

\begin{figure}
\centering
\epsfig{file=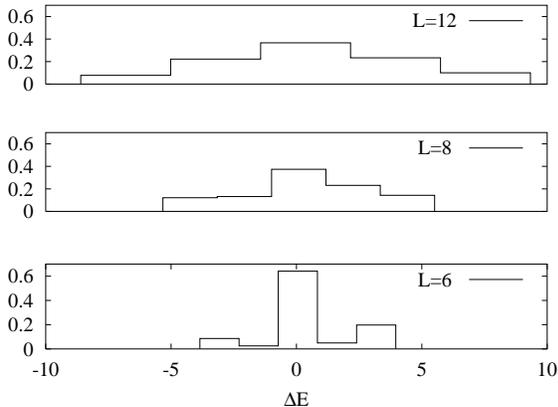,width=0.45\textwidth}
 \caption{Probability distribution of $\Delta E$, for $L=6$, $8$, and $12$.}
\label{fig_P_DeltaE}
\end{figure}

Second, the spread of the distribution of $\Delta E$ grows with $L$;
this is very visible in figure~\ref{fig_P_DeltaE} and is to be
contrasted with the behavior of $\Delta F$. To make this point more
quantitative, we have calculated the variances of $\Delta E$; these
are given in table~\ref{tab_Energy_fluctuations}, confirming that the
distribution of $\Delta E$ broadens as $L$ increases.  Third, we find
that the linear correlation coefficient of $\Delta E$ and $\Delta F$
is small.  This coefficient is defined as

\begin{equation}
C(\Delta E, \Delta F) =  {\frac
{\overline {(\Delta E - \overline{\Delta E})(\Delta F - \overline{\Delta F})}}
{\sigma (\Delta E) ~~ \sigma (\Delta F) } }\ ,
\end{equation}
where $\sigma$ is the standard deviation of the variable of interest
and an over-bar stands for the disorder average. The values of $C$ are
also given in the table.  It would be nice to test the possibility
that the variance of $\Delta E$ grows as a power of $L$,
but our range of lattice sizes is too small to
seriously test this hypothesis. Nevertheless, the picture we extract
from our analysis is that these ``large'' fluctuations in energy must
be compensated by similar fluctuations in entropy, otherwise one could
not have $\Delta F = O(1)$. Most probably, these fluctuations play a
crucial role in all frustrated disordered systems and they deserve
further investigation.

\begin{table}
\begin{center}
\caption{$\Delta F$, mean and variance of $\Delta E$, and correlation 
coefficient of $\Delta E$ and $\Delta F$.}
\label{tab_Energy_fluctuations}
\begin{tabular}{lllll}
\hline\noalign{\smallskip}
L & $\overline {\Delta F}$ & $\overline {\Delta E}$ & $\sigma(\Delta
E)$ & $C(\Delta E, \Delta F)$\\
\noalign{\smallskip}\hline\noalign{\smallskip}
6  & 0.317 & 0.418 & 1.879 & 0.048 \\
8  & 0.352 & 0.364 & 2.409 & 0.005 \\
12 & 0.328 & 0.771 & 4.123 & 0.114 \\
\noalign{\smallskip}\hline
\end{tabular}
\end{center}
\end{table}

\section{Inter-valley differences: spin and link overlaps}
\label{sect_spin_link_overlaps}

We focus here on the relation between spin and link overlaps: are
these two fluctuating quantities correlated? A motivation for this
question comes from mean field theory where $q_l$ becomes a {\it
deterministic} function of $q$ in the thermodynamic limit. To
investigate this possibility, we look at the variance of $q_l$ at
fixed $q$.  There are several approaches: the first considers all
fluctuations; the second focuses on instance to instance fluctuations
only; the third is based on the fluctuations given by the TAP-like
states of our valleys.

Consider the first approach. We take equilibrium configurations at
random, and determine (for $q$ in a small range) the distribution of
$q_l$ for $L=6$, $8$, and $12$.  This distribution is averaged over
the disorder, and we consider the resulting variance. We find that the
variance decreases as one increases $L$; for example, in the bin $0
\le q \le 0.1$, the standard deviation is $0.043$ at $L=6$, $0.033$ at
$L=8$, and $0.028$ at $L=12$. (The other bins give very similar
results.) This decrease is unambiguous, and is highly suggestive of an
asymptotic decrease to $0$ as $L \to \infty$.  Previous
work~\cite{MarinariParisi00c} on ground states lead to this same
conclusion.

Fluctuations in $q_l$ come from thermal noise within a given instance
{\it and} from instance to instance fluctuations.  In a na\"{\i}ve
way, assume that the variances add linearly, $\sigma^2(q_l) \approx
\sigma_T^2(q_l) + \sigma_J^2(q_l)$ where the first one is the
intra-instance (purely thermal) variance and the second one is the
inter-instance variance.  The thermal fluctuations are due to
thermally excited droplets inside the valleys; these fluctuations are
expected to average out in any scenario, so that
$\sigma_T^2(q_l) \to 0$ at large $L$. The measurements of the previous
paragraph show that $\sigma^2(q_l)$ decreases as $L$ grows, but this
can happen even if $\sigma_J^2(q_l)$ saturates at a positive value. It
is thus appropriate to determine the $L$ dependence of
$\sigma_J^2(q_l)$ by itself.  This leads us to our second method
whereby we find the mean of $q_l$ in a given disorder instance (and
for $q$ in a given bin) and then we consider its instance to instance
fluctuations.  Naturally, the variances now are smaller than in the
previous measurements. Is there a sign of saturation?  Let us again
give our results for $0 \le q \le 0.1$: at $L=6$,
$\sigma_J(q_l)=0.034$, at $L=8$, $\sigma_J(q_l)=0.028$, and at
$L=12$, $\sigma_J(q_l)=0.022$. We have the same decreasing trend as
before, strengthening the claim that $q$ and $q_l$ are related
deterministically at large $L$.

In the second method just described, we removed the thermal noise in
$q_l$; can one also remove the thermal noise in $q$? Clearly this can
be meaningful only if we restrict the averaging so that one stays in
given valleys. A simple approach is then to take the mean of $q$ for
all pairs of configurations belonging to two different 
but fixed valleys. It is
not difficult to see that the corresponding mean overlap is equal to
the overlap of the TAP-like states of the two valleys. Let these
states be ${\cal M}^{(1)}$ and ${\cal M}^{(2)}$, having magnetizations
$\{ m_i^{(1)} \}$ and $\{ m_i^{(2)} \}$; the valley-to-valley spin
overlaps average to

\begin{equation}
q({\cal M}^{(1)},{\cal M}^{(2)}) \equiv
{\frac {1}{N}} {\sum_i m_i^{(1)} m_i^{(2)} }
\end{equation}
exactly. One is then tempted to extend the analysis by considering
the link overlap of these states
\begin{equation}
q_l({\cal M}^{(1)},{\cal M}^{(2)}) \equiv
{\frac {\sum_{\langle ij\rangle} m_i^{(1)} m_i^{(2)} m_j^{(1)} m_j^{(2)}
}{\sum_{\langle ij\rangle} 1}}\
\end{equation}
and seeing whether these two overlaps are related deterministically at
large $L$. We believe that such a test 
would be appropriate for systems with one-step RSB. But here we have
continuous RSB so our TAP-like states are associated with valleys that
can have sub-valleys; then, even if there is a deterministic relation
between $q$ and $q_l$ at the level of individual configurations, that
will no longer be the case when considering $q({\cal M}^{(1)},{\cal
M}^{(2)})$ and $q_l({\cal M}^{(1)},{\cal M}^{(2)})$. Indeed, the
averaging over the sub-valleys introduces intrinsic fluctuations
that will not go to $0$ as $L \to \infty$. So instead, let us look at
the problem differently. To remove the thermal fluctuations, we should
remove the droplet excitations.  Suppose we think of a valley as a
``reference configuration'' dressed by any number of droplet
excitations. Granted, this picture is simple-minded,
but it provides a useful framework.  When computing the TAP-like state
of that valley, we start with the reference configuration and let the
thermal fluctuations decrease its magnetizations from $\pm 1$. This
suggests that if we bring each $m_i$ back to its original value, we
will reconstruct this reference configuration which has no thermal
noise in it.  Our procedure for approximating this reference
configuration is simply to project each $m_i$ of the TAP-like state
back to $\pm 1$ according to its sign; we call the resulting
configuration the ``projected TAP-like state''. It plays the role of a
reference configuration, a kind of ``heart'' of the valley.

In figure~\ref{fig_scatter} we show the collection of spin and link
overlaps among these projected TAP-like states in our data sample.
The data are scattered, but there is also a clear trend.  In
figure~\ref{fig_scatter_fit} we show the averaged data, restricted to
$L=12$, with binning in $q$.  Superposed in this figure is the best
quadratic fit to the data (a quadratic dependence is what one has in
the mean field theory). The fit is very satisfactory.

Two questions of interest are: 
\begin{enumerate}

\item What is the limiting shape of this curve at large $L$?

\item In fig.~\ref{fig_scatter}, does the scatter go to $0$ when $L$ grows?

\end{enumerate}
For the limiting curve, we expect it to be a monotonic function, $\langle q_l
\rangle$ growing with $q$. For conciseness, let us then just give the
values for $\langle q_l \rangle$ at $q=0$. To estimate this mean, we
consider all events in the window $0 \le q \le 0.1$, leading to
$\langle q_l \rangle$ equal to $0.428 \pm 0.007$ at $L=6$, $0.483 \pm
0.005$ at $L=8$, and $0.555 \pm 0.004$ at $L=12$. The trend is towards
increasing values with $L$, just as was found for low-energy
excitations above the ground
state~\cite{KrzakalaMartin00a,PalassiniYoung00a,MarinariParisi00c,HoudayerKrzakala00}
and from Monte Carlo simulations~\cite{KatzgraberPalassini00}.
Note also that the numbers themselves are very close to those
obtained by Houdayer {\it et al.}~\cite{HoudayerKrzakala00}.

\begin{figure}
\centering
\epsfig{file=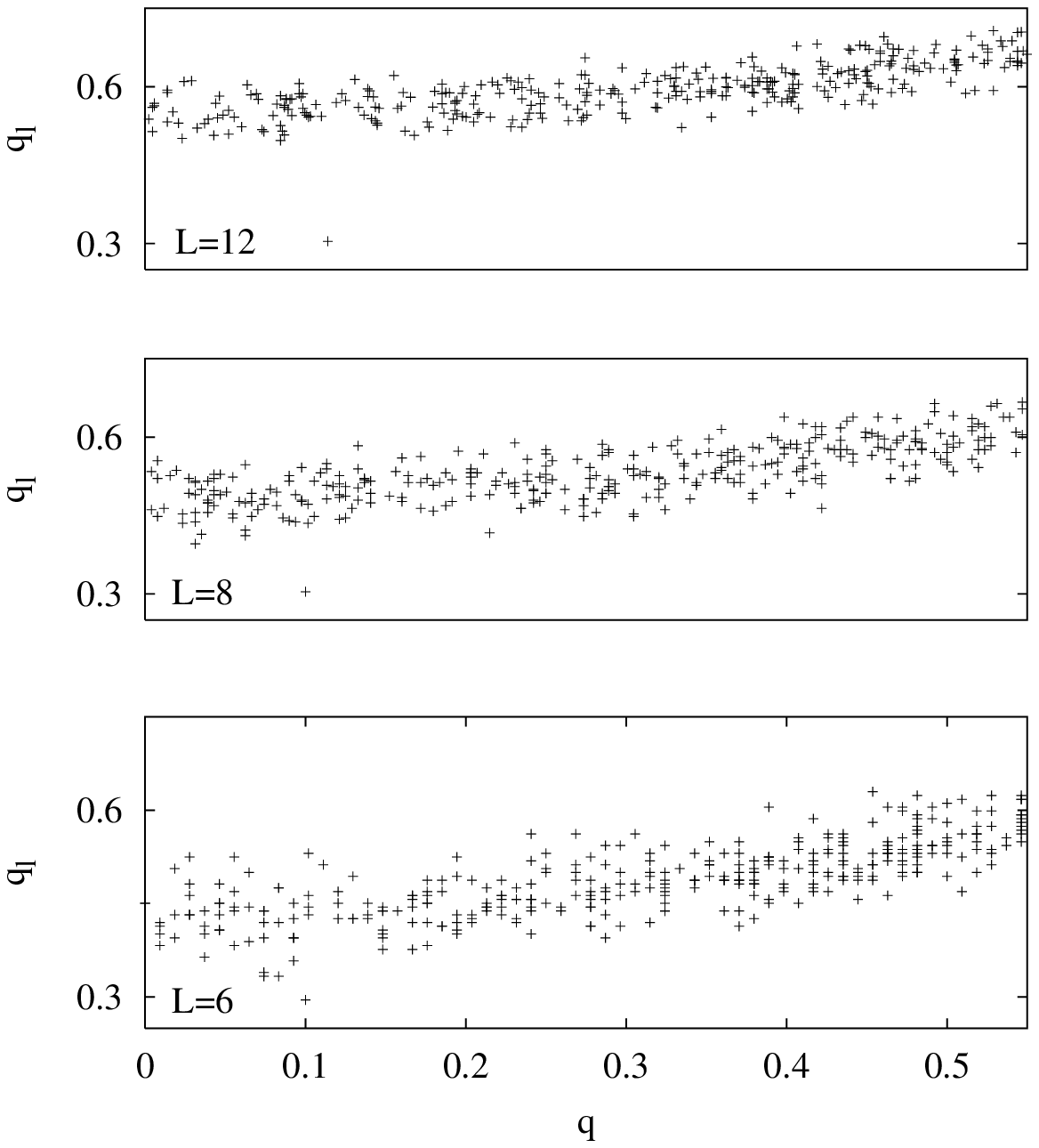,width=0.45\textwidth}
 \caption{Scatter plot of $q_l$ versus $q$ for the projected TAP-like 
states.}
\label{fig_scatter}
\end{figure}

\begin{figure}
\centering
\epsfig{file=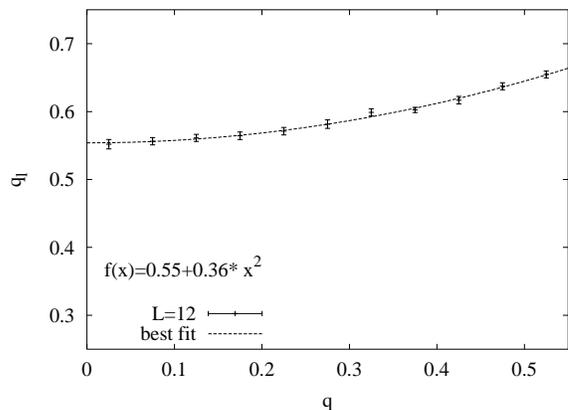,width=0.45\textwidth}
 \caption{Disorder averages of $q_l$ at fixed (binned) $q$ values
	  where $q$ and $q_l$ are for the projected TAP-like
	  states. The dashed curve is the best fit to a quadratic dependence.}
\label{fig_scatter_fit}
\end{figure}
To address the second question, we have measured the variance of $q_l$
when $q$ is restricted to a given range.  As above, consider the
window $0 \le q \le 0.1$; in this range, the standard deviation of
$q_l$ is $0.045$ at $L=6$, $0.037$ at $L=8$, and $0.031$ at
$L=12$. These variances decrease with $L$. Again, our range in $L$ is
too small for us to meaningfully fit these values to a 
constant plus power law in $1/L$. The main point is that the trend is
consistent with what was found previously at zero and finite temperature,
and gives one confidence that in the large $L$ limit $q$ and $q_l$
are related by a deterministic law.

\section{Inter-valley differences: position space picture}
\label{sect_position_space_picture}

In our working conditions, {\it i.e.}, at quite low $T$, we find that
many of the magnetizations $m_i$ defining the TAP-like states are
close to $\pm 1$.  This suggests that the projection to $\pm 1$ is a
reasonable method to reach a valley's heart. Furthermore, 
with this projection, we
can compare with previous results at $T=0$.  At stake are the spatial
properties of the differences of projected TAP-like states.

To motivate our measurements, let us first ``see'' what these
differences actually look like. In figure~\ref{fig_sponge} we show an
example of a difference (called $G$) at $L=12$ for an instance that
passed the selection tests described previously.

\begin{figure}
\centering
\epsfig{file=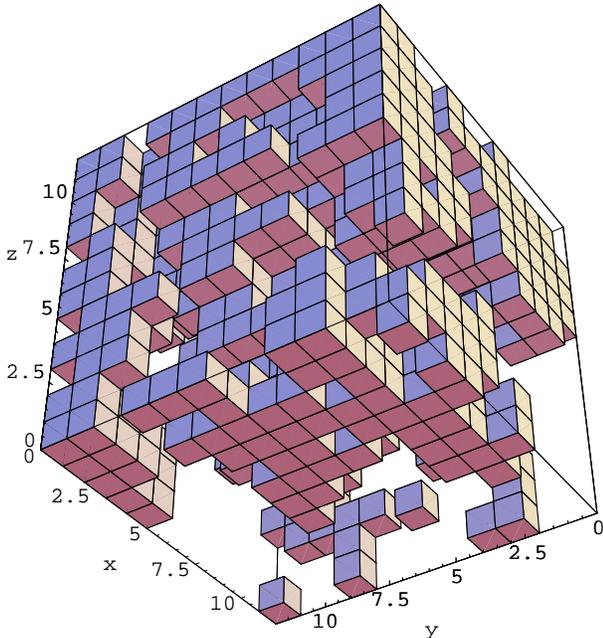,width=0.45\textwidth}
 \caption{Difference of two pro\-jec\-ted TAP-li\-ke states ($L=12$ lattice).}
\label{fig_sponge}
\end{figure}
The cluster is very space filling; to make this clear, we recommend
looking at the display on a computer screen while it is being drawn
(this gives far more information than the final printed picture where
most of the cluster is hidden). Furthermore, the cluster is
topologically non-trivial: one can guess at the numerous handles in
its inside, and not surprisingly in our analysis it turns out to be
sponge-like (it winds around the lattice in the three directions). One
of the central questions for these types of objects concerns the scale
beyond which (if any) they become homogeneous. This is a difficult
question so it is wisest to focus on the cases for which $q$ is small:
if any kind of homogeneity can be detected, it should be testable
there.

As a probe of heterogeneities, we follow Houdayer et al.~\cite{HoudayerKrzakala00} and
ask whether it is possible to push a needle all the way through the
lattice without hitting the boundary of $G$ (the difference of the two
projected TAP-like states). This has a direct and intuitive
interpretation: let each site of $G$ be associated with a unit cube
that is opaque, as displayed in the figure. We send light onto one
face of the cube and ask whether $G$ stops all the light or whether
some light is able to pass through the cube without being absorbed. If
$G$ is homogeneous on all scales larger than the
lattice spacing, the probability of absorbing all the light goes to
$1$ as $L\to \infty$. Because of the up-down symmetry, we ask in fact
whether both $G$ and its complement absorb all the light. We find, as
in the $T=0$ landscape studies \cite{HoudayerKrzakala00}, that there
is almost always a straight line parallel to a given axis that stays
entirely in $G$ or entirely in its complement.  The system is thus not
opaque. We can extend this kind of study by generalizing the needle to
a pencil or ``tube'': now we try to draw not a single line, but $4$
parallel lines whose cross-section is a plaquette of our lattice (the
basic square formed by $4$ bonds connecting nearest neighbors). The
fatter the tube, the less likely it will be possible to slide it
through without hitting the surface of $G$. To compare with the $T=0$
study, we have used the same cross-sections as there: tube ``2'' has the
plaquette cross-section with $4$ sites; tube ``3'' has a diamond
cross-section with $5$ sites, i.e., a central site and its $4$ nearest
neighbors; finally tube ``4'' has for its cross section two adjacent
plaquettes with a total of $6$ sites (the reader can find
pictures of these cross-sections in Houdayer et al.~\cite{HoudayerKrzakala00}.)

Table~\ref{tab_spanning_tube} gives the fraction of events for which
there is at least one tube along the $x$ axis which ``spans'' the
whole cube (i.e., is of length $L$) without touching the surface of
$G$ (the tube sites are thus entirely in $G$ or in its complement).
The data are for those disorder instances with at least two 
valleys and with the
extra constraint that $0 \le q \le 0.1$ where $q$
is the spin overlap of the projected TAP-like states.

\begin{table}
\begin{center}
\caption{Fraction of events allowing a spanning tube
through $G$ as a function of the tube cross-section type.}
\label{tab_spanning_tube}
\begin{tabular}{llll}
\hline\noalign{\smallskip}
L & Tube ``2'' & Tube ``3'' & Tube ``4'' \\
\noalign{\smallskip}\hline\noalign{\smallskip}
6  & 0.44 & 0.18 & 0.09 \\
8  & 0.49 & 0.13 & 0.20 \\
12 & 0.46 & 0.19 & 0.16 \\
\noalign{\smallskip}\hline
\end{tabular}
\end{center}
\end{table}

These data are a bit too noisy to extract a reliable trend. At best, we
can say that the data are compatible with increasing values of the
spanning probabilities as $L$ grows; that is the behavior arising in
the energy landscape study.  This would mean that the difference of
two projected TAP-like states is geometrically similar to the difference of
two low-lying system-size excited states.
These findings should be considered together with the
evidence that the clusters are space-filling
(see~\cite{MarinariParisi00a,MarinariParisi00b}): a $2D$ plane will
intersect the clusters with very high probability. Thus on scales
comparable to $L$, the clusters are space filling, in agreement with
the Parisi mean field theory.

\section{Discussion and conclusions}
\label{sect_conclusions}

We believe that the analysis presented in this note sheds significant
light on the issue of the behavior of finite dimensional, realistic
spin glasses. On the one hand we stress the physical relevance of the
picture of space filling sponges; on the other hand our 
numerical results from low $T$ Monte Carlo simulations give evidence for RSB
with a mean field like scenario. Perhaps our most important point is
that the results of low $T$ numerical simulations give the same
indications as the ground state computations. This fact attenuates the
long-standing doubt: ``Are spin glass numerical simulations
representative of low $T$ physics, or do they just contain artifacts
due to the $T_c$ critical point?'' Our answer is then that indeed
recent numerical simulations {\it do} give information about the low
$T$ physics. Similarly, ground state computations then characterize the low $T$
physics behavior of these systems.

A second important issue we address concerns entropic effects. We
find that the mean field like features of the system arise with
entropy and energy fluctuations; the mutual cancellation of these
fluctuations allows the near degeneracy in the
free-energy of different states. To see these fluctuations, it is
necessary to introduce valleys that are in a sense the building blocks
of what can be called finite volume equilibrium states.

A third point worth mentioning is that definitions such as sponge-like
or droplet-like differences are very useful; not only do they seem
more and more clear-cut as $L$ grows, but they also allow for a finer
analysis.  An illustration of this potential was given via a modified
spin overlap (restricted to sponges) that appears to be more robust than
the usual one. For the future, we expect these definitions to be
extremely useful when trying to understand putative deviations of
realistic spin glasses from the standard mean field picture.

There is a lot more that can be done; many issues 
should be understood in greater detail and checked on larger lattices
(our largest size is $L=12$). Let us just give a bird's-eye
view of what we feel are some of the main questions. (1): All of our
work was for a given temperature; the effect of raising and lowering
the temperature should be investigated.  (2): The study of the
structure of valleys has only begun. Our focus here was on large and
well separated valleys; clearly a complete picture requires one to
consider the more general case.  (3): We have analyzed the EA model
with binary ($\pm$) couplings. For the lattice sizes we used, this
discreteness showed up on the mean energy differences of valleys (see
figure~\ref{fig_P_DeltaE}). Does this effect go away for large sizes and
can one confirm numerically the expectation that the Gaussian
couplings lead to similar results at large $L$?

\vspace{1cm}
\centerline{\bf Acknowledgments} 

\bigskip
We thank A. Billoire, J.-P. Bouchaud, E. Domany, J. Houdayer, F. Krzakala,
M. M\'ezard and G. Parisi for very stimulating discussions. The LPTMS
is an Unit\'e de Recherche de l'Universit\'e Paris~XI associ\'ee au
CNRS.  One of us (EM) warmly acknowledges the hospitality of
the LPTMS during which part of this work was done.
The numerical simulations were run on {\em Kalix2}, a Linux
cluster located at the Physics Department of Cagliari University, and
funded by MURST COFIN 1998 (Italy). FZ is supported by an EEC
Marie Curie fellowship (contract HPMFCT-2000-00553).

\vspace{1cm}
\centerline{{\bf Appendix}}

\bigskip
We want to characterize {\em valleys} or families
of spin configurations by their
local magnetizations, ``\`a la TAP'', in analogy with what occurs in
mean field models. However, from a thermodynamical point of view, it
is not easy to define TAP states in a finite volume. Because of this,
our definition of valleys is obtained by clustering according to
the spin overlaps. One can consider the formal approach of clustering
over all of configuration space where each configuration has a given
probability, or one can use a more ``hands-on'' description, where one
clusters equilibrium configurations (as obtained from our
simulation). Since it is simpler to explain our clustering in the
latter framework, we limit ourselves to that case here.

To cluster our configurations, we will use the mutual
overlaps and a cut-off parameter $q^*$. To begin, we give two
definitions.  A valley is just a set of configurations; these will be
selected by the clustering algorithm, but what matters is that for
each valley $k$ one has a collection of configurations ${\cal C}_n^k$
($n=1, ... |V_k|$).  Second, the TAP-like state associated with valley
$k$ is the list of magnetizations $m_i^{(k)} \in [-1,1]$, one for
each site $i$ of the lattice.  By definition, we take $m_i^{(k)}$ to
be the mean of $S_i$ when considering all the configurations belonging
to valley $k$.

The logic of our approach can be summarized as follows: (i) a valley
defines its TAP-like state; (ii) the TAP-like states tell us whether
or not a new configuration should be assigned to one of the $k$
valleys. A sensible assignment rule is to impose that the overlap of
this new configuration with the corresponding TAP-like state should be
sufficiently large. In our algorithm, we first compute the spin
overlaps between the new configuration and the $k$ TAP-like states;
let $q^{(1)}$, $q^{(2)}$, ... $q^{(k)}$ be these overlaps. The new
configuration is then assigned to valley $j$ if $q^{(j)}$ is the
largest of these overlaps {\it and} if $q^{(j)} \ge q^*$.

With these definitions, we can now explain our clustering procedure
which is based on defining valleys and their TAP-like states {\it
self-consistently}.  Suppose we have built $k$ valleys so far; let
their TAP-like states be ${\cal M}^{(1)}$, ${\cal M}^{(2)}$, ...,
${\cal M}^{(k)}$. Suppose we want to increase $k$ by $1$. First, we
initialize ${\cal M}^{(k+1)}$ by setting its $m_i^{(k+1)}$
to be equal to spin values $S_i$ of one of the
configurations that is not assigned to any valley.  Then we go through
the rest of the un-assigned configurations, adding them successively
to the pool defining the ($k+1$)th valley if their overlap $q^{(k+1)}$
with ${\cal M}^{(k+1)}$ is greater than $q^*$ (the overlap is defined
in the usual way, $q = \sum_1^N S_i m_i / N$).  Every time a
configuration is added to the pool, we update ${\cal M}^{(k+1)}$ so
that it always gives the average magnetization at each site for the
configurations in valley $k+1$.

At the end of this pass, we have ${\cal M}^{(1)}$, ${\cal M}^{(2)}$,
..., ${\cal M}^{(k+1)}$. But because ${\cal M}^{(k+1)}$ changes as one
adds configurations to the pool, some of such configurations may see
their overlap $q^{(k+1)}$ go below $q^*$. Furthermore, some of the
configurations assigned to valleys $j$ ($j \le k$) may end up being
closer to ${\cal M}^{(k+1)}$ than to ${\cal M}^{(j)}$. It is thus
necessary to rebuild all the valleys, so we first de-assign all the
configurations and then we perform a new pass where a configuration is
assigned to valley $r$ if both $q^{(r)} > q^*$ and $q^{(r)} > q^{(s)}$
for all $s \ne r$. After three iterations of this build/un-build
process, our algorithm takes these to be the valleys and computes the
corresponding ($k+1$) TAP-like states. These additional iterations
improve the self-consistency and reduce the valleys'
dependencies on the ordering of the configurations.
This procedure is very close in spirit to that given by Iba and
Hukushima~\cite{IbaHukushima00}; they also have TAP-like states and
valleys that are defined self-consistently by an iterative algorithm.

The last issue pertinent to our method concerns the total number $M$
of valleys to be generated.
In our implementation, we continue creating more valleys until either
of the three following criteria are met: (1) all
configurations are assigned to valleys; (2) one has reached
the maximum number of valleys (set to $10$ in our code); 
(3) adding one more valley and its TAP-like state would lead
to ($M+1$) TAP-like states whose cross-overlaps would not all be
less than $q^*$. When that happens, we consider that a valley
has been subdivided too far and we do not accept this
($M+1$)th TAP-like state.

\bibliographystyle{prsty}
\enlargethispage{30pt}
\bibliography{references}

\end{document}